\documentclass[aps,prl,twocolumn,superscriptaddress,showpacs]{revtex4}

\usepackage{graphicx, latexsym, verbatim}
\usepackage{amssymb, amsmath}
\usepackage[ansinew]{inputenc}
\usepackage{color}
\usepackage[english]{babel}

\begin{document}

\title{Quantifying Transition Voltage Spectroscopy of Molecular Junctions}

\author{Jingzhe Chen}
\affiliation{Center for Atomic-scale Materials Design (CAMD), Department of Physics,
Technical University of Denmark, DK-2800 
Kgs. Lyngby, Denmark}

\author{Troels Markussen}
\email[]{trma@fysik.dtu.dk}
\affiliation{Center for Atomic-scale Materials Design (CAMD), Department of Physics,
Technical University of Denmark, DK-2800 Kgs. Lyngby, Denmark}
\affiliation{Danish National Research Foundations Center of Individual Nanoparticle Functionality (CINF), Department of Physics, Technical University of Denmark, DK-2800 Kgs. Lyngby, Denmark}

\author{Kristian S. Thygesen}
\affiliation{Center for Atomic-scale Materials Design (CAMD), Department of Physics,
Technical University of Denmark, DK-2800 Kgs. Lyngby, Denmark}

\date{\today}

\begin{abstract}
  Transition voltage spectroscopy (TVS) has recently been introduced
  as a spectroscopic tool for molecular junctions where it offers the
  possibility to probe molecular level energies at relatively low bias
  voltages. In this work we perform extensive \textit{ab-initio}
  calculations of the non-linear current voltage relations for a broad class
  of single-molecule transport junctions in order to assess the
  applicability and limitations of TVS. We find, that in order to fully
  utilize TVS as a quantitative spectroscopic tool, it is important to
  consider asymmetries in the coupling of the molecule to the two
  electrodes. When this is taken properly into account, the relation
  between the transition voltage and the energy of the molecular
  orbital closest to the Fermi level closely follows the trend
  expected from a simple, analytical model.
\end{abstract}


\maketitle
Molecular electronics holds the promise of continuing the
miniaturization of electronic devices beyond the limits of standard
silicon technologies by using single molecules as the active
elements~\cite{poulsen09,SongNature2009}. In the design and
characterization of molecular devices, the electronic structure of the
molecule is naturally of vital importance. In fact, when quantum
interfererence effects are disregarded\cite{baer_neuhauser02}, there is a direct
relation between a junction's transport properties and the distance of
the molecular energy levels to the electrode Fermi level. The latter
may in principle be determined from peaks in the $\text{d}I/\text{d}V$ curve, where
$I$ is the current and $V$ is the bias voltage. Assuming that the
molecular level closest to the Fermi level is the highest occupied
molecular orbital (HOMO) it would require a voltage of
$V \sim 2|E_F-\varepsilon_H|$ to probe the HOMO
position in a symmetric junction. However, in practice the molecular junction often becomes
unstable and breaks down due to the large current density before the
peak in the $\text{d}I/\text{d}V$ is reached.

Transition voltage spectroscopy (TVS) was introduced as a
spectroscopic tool in molecular electronics by Beebe \textit{et al.}
~\cite{Beebe2006}. Beebe found that the Fowler-Nordheim graph of a
molecular junction, i.e. a plot of $\ln(I/V^2)$ against $1/V$, showed
a characteristic minimum at a bias voltage $V_{\rm min}$ which scaled
linearly with the HOMO energy obtained from ultraviolet photo
spectroscopy (UPS). Importantly, the TVS minimum is obtained at
relatively low bias voltage before electrical break-down. TVS is now
becoming an increasingly popular tool in molecular
electronics~\cite{Noguchi2007,Beebe2008,Zangmeister2008,HoChoi2008,SongNature2009,TanAPL2010}.

The original interpretation of TVS introduced by Beebe~\cite{Beebe2006} and applied
in most later experimental work, is based on a Simmons tunnel barrier
model~\cite{Simmons}. Within this interpretation, the TVS minimum is obtained, when
the tunnel barrier, due to the applied bias potential, changes from being
trapezoidal to triangular. The transition voltage equals the barrier height, which
is interpreted as the distance from the Fermi level to the closest molecular level.
However, it has recently been pointed out by Huisman {\it et al.} that the barrier
model is inconsistent with experimental data \cite{Huisman2009} which on the other hand is more appropriately described by transport via a single electronic level. In the single-level model the transmission function was assumed to have a Lorentzian shaped, and all non-linear effects due to the finite bias, were neglected. While these
assumptions may be reasonable, they are not obviously fulfilled in a realistic
molecular junction at high electric fields. It is therefore of interest to compare
the simple Lorentzian transmission model with more realistic calculations. 

In this Letter, we present a quantitative analysis of TVS based on extensive  
\emph{ab-initio} calculations of the non-linear current-voltage relations for a broad
class of molecular junctions. The ratio of the
TVS minimum to the HOMO level position is found to vary between 0.8 and 2.0
depending on the junction asymmetry, i.e.
quite different from the one-to-one relation assumed so far. The large
variation is due to the difference in the non-linear response of the
molecular level to the bias voltage.  The importance of asymmetry effects is
further signified by the fact that many of the experiments using TVS
were performed on asymmetric molecules in an asymmetric conductive AFM
measurement
setup~\cite{Beebe2006,Beebe2008,Zangmeister2008,TanAPL2010}. Indeed,
as we show below a larger degree of consistency between TVS
measurements and UPS data is obtained when the asymmetry is taken into
account.

\begin{figure}[htb!]
\includegraphics[width=0.45\columnwidth]{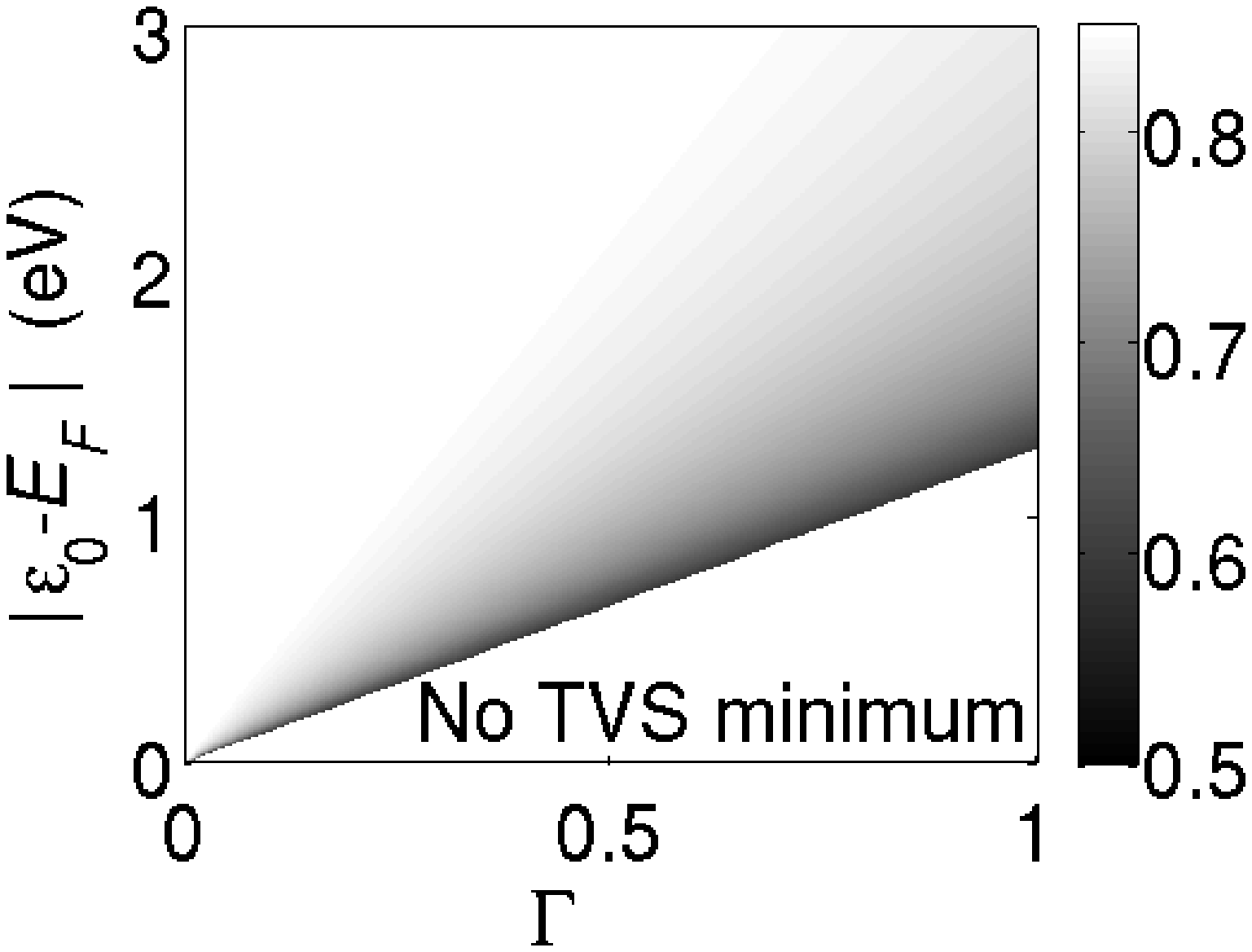}
\put(-87,80){$|\varepsilon_0-E_F|/V_{\rm min}$}
\put(-62,67){$\eta=0$}
\includegraphics[width=0.45\columnwidth]{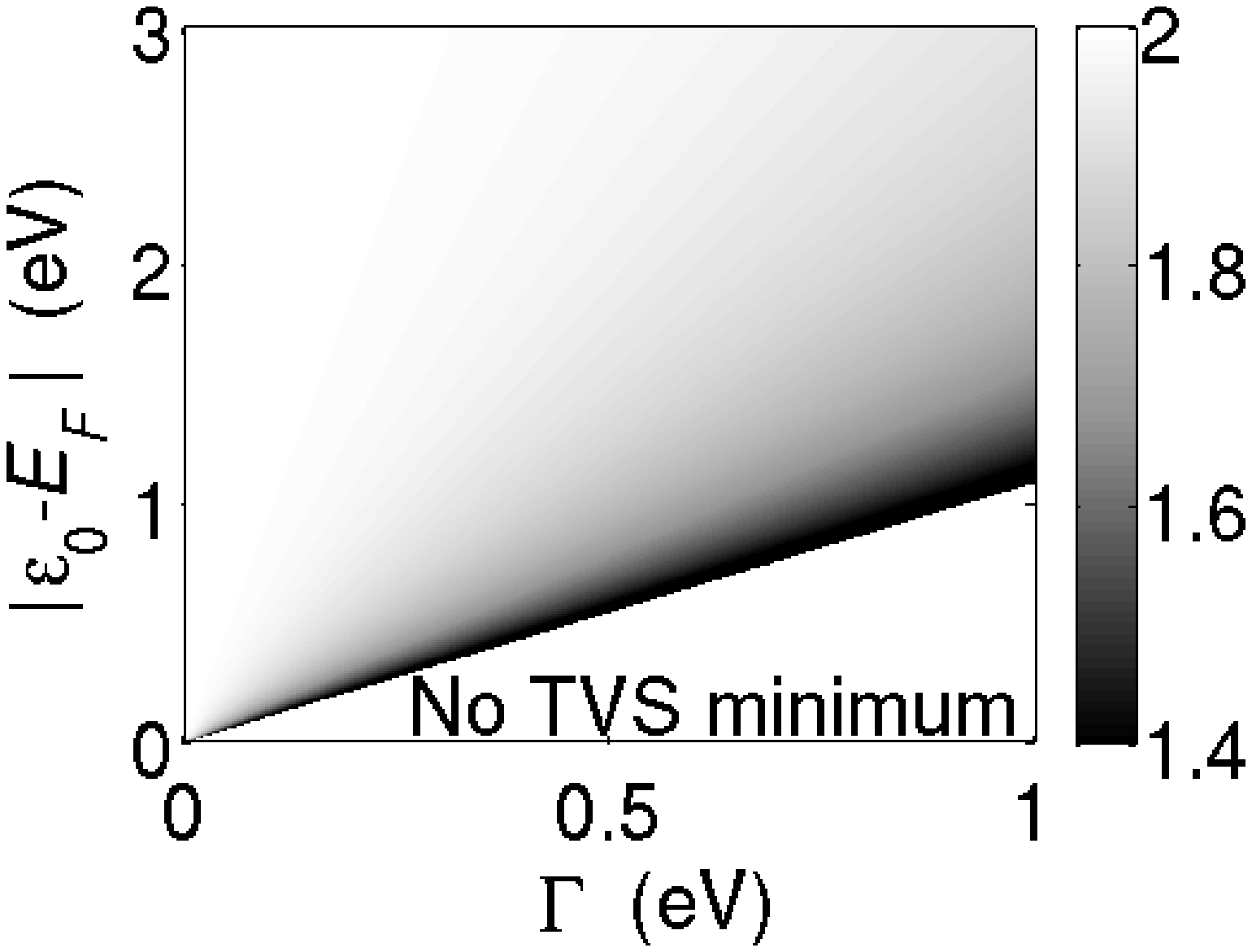}
\put(-87,80){$|\varepsilon_0-E_F|/V_{\rm min}$}
\put(-62,67){$\eta=1/2$}

\includegraphics[width=0.56\columnwidth]{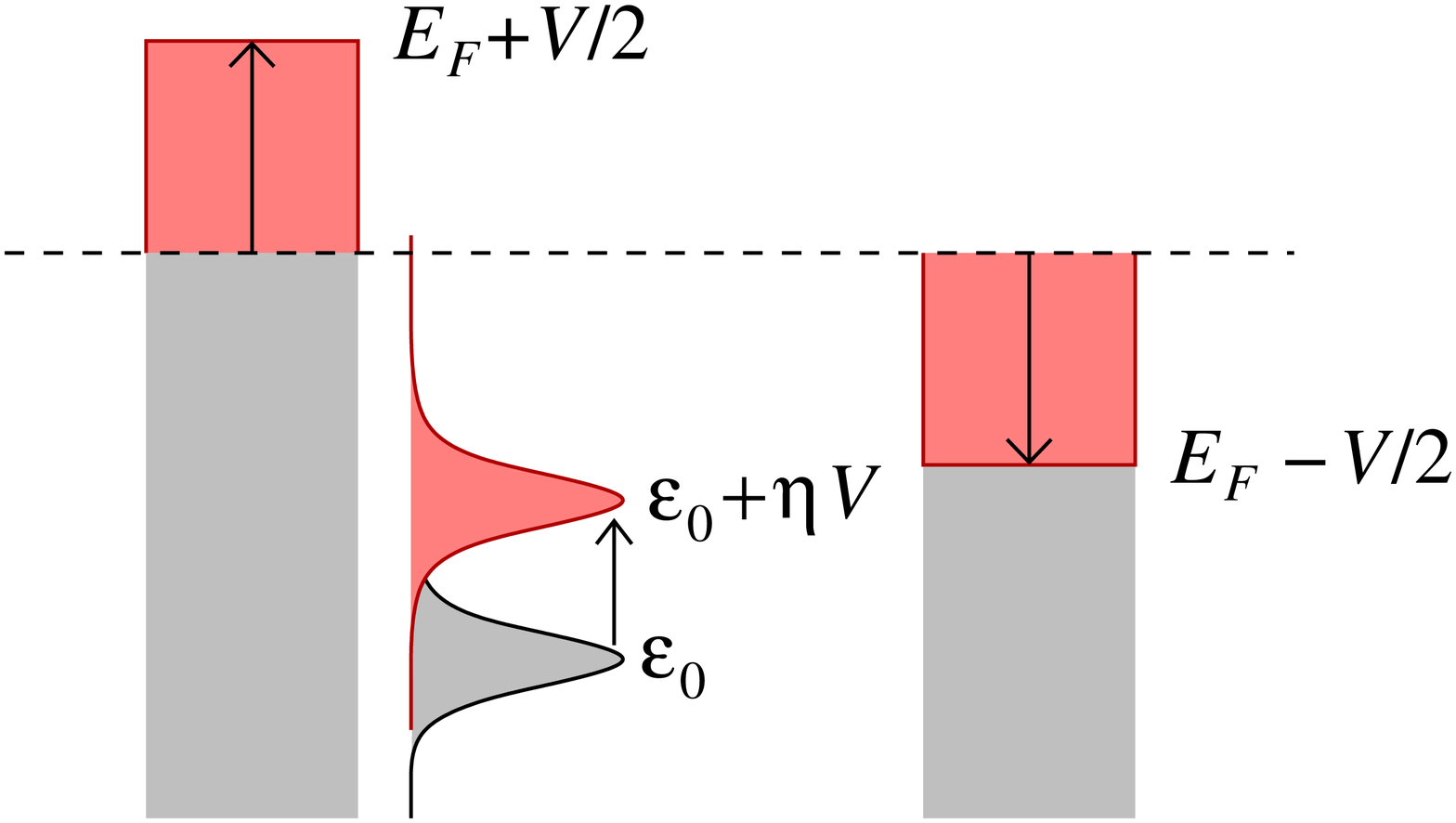}
    \caption{Ratio between molecular level energy and transition voltage,
$|\varepsilon_0-E_F|/V_{\rm min}$, vs. molecular level energy, $\varepsilon_0$ and
broadening, $\Gamma$, for a symmetric junction, $\eta=0$, (left), and a
completely asymmetric junction, $\eta=1/2$, (right). Note the different scales of $|\varepsilon_0-E_F|/V_{\rm
min}$ for the symmetric and asymmetric junctions. The lower part illustrates how the molecular level moves under
a finite bias voltage. At zero bias the level is located at $\varepsilon_0$, but
when a bias voltage is applied, the level follows to some degree the chemical
potential of the left electrode, due to a stronger coupling to this electrode.}
\label{tvs_analytical}
\end{figure}

We begin our analysis by considering a simple model for transport via a single electronic level equivalent to the work in Ref. \cite{Huisman2009}. The transmission
function is assumed to be a Lorentzian
\begin{equation}
 T(E;\varepsilon_0,\Gamma) = \frac{f}{(E-\varepsilon_0)^2+\Gamma^2/4},
\label{Transmission}
\end{equation}
where $\varepsilon_0$ and $\Gamma$ are the molecular level energy and broadening,
respectively. $f$ is a constant factor to account for multiple molecules in the
junction, asymmetric coupling to the electrodes, etc. Our analysis will not be
dependent on the actual value of $f$. We shall assume that the molecular level is the HOMO level, i.e. $\varepsilon_0<E_F$. This is the typical case for thiol bonded molecules (as considered in this work), where electron transfer from the metal to the sulphur end group shifts the molecular levels upward in energy\cite{guowen09}.
At finite bias voltage, the molecular level may be shifted relative to the zero-bias position, as shown schematically in Fig.
\ref{tvs_analytical} (lower part). Such non-linear effects express themselves differently in symmetric and asymmetric junctions \cite{KerguerisPRB1999,SenseJanReview}. For example, if the molecule is only strongly coupled to the left electrode, the molecular levels will follow the chemical potential of the left contact. On the other hand, for a symmetric junction the molecular level will remain at the zero-bias position. We describe the degree of asymmetry by the parameter
$\eta\in[-1/2; 1/2]$ such that the current is given by
\begin{equation}
I = \frac{2e}{h}\int_{-\infty}^{\infty}T(E;\varepsilon_0+\eta
\,V,\Gamma)\left[f_L(V)-f_R(V) \right]{\rm d}E, \label{Current}
\end{equation}
where $f_{L/R}(V)=1/[\exp{(E_F\pm e V/2)/k_B T}+1]$ are the
Fermi-Dirac distributions for the left and right contact,
respectively. The dependence of the level position on
the bias voltage is the main difference between our model and the one
considered in Ref. \onlinecite{Huisman2009}.

Using Eq. \eqref{Current} we calculate $\ln(I/V^2)$ and find the minimum at the bias
voltage $V_{\rm min}$. At this voltage, the slope of the current depends
quadratically on the bias voltage, $I\propto V^2$, and $\text{d}(\ln(I/V^2))/\text{d}V=0$. An
example of a calculated Fowler-Nordheim plot can be seen in the inset of Fig.~\ref{TVS_pdos_C6}. In Fig. \ref{tvs_analytical} we plot the ratio $|\varepsilon_0-E_F| / V_{\rm min}$ as a function of broadening, $\Gamma$ and molecular level
energy, $\varepsilon_0$, for a symmetric junction ($\eta=0$) and a completely asymmetric
junction ($\eta=1/2$). Fig. \ref{tvs_analytical} illustrates two main points: First, the
ratio $|\varepsilon_0-E_F|/V_{\rm min}$ is nearly constant over a large range of $\Gamma$
and $\varepsilon_0$ values. For a given degree of asymmetry (value of $\eta$), the TVS minimum can therefore be used as a direct measure of the molecular level position, independently of $\Gamma$. Note, however, that for $\Gamma/|\varepsilon_0-E_F| \gtrsim 1$, there is no minimum in the Fowler-Nordheim plot, since the
molecular level is too close to the Fermi level, and the current increases slower than $\propto V^2$ at all bias values. The second conclusion from Fig. \ref{tvs_analytical} is
that the ratio $|\varepsilon_0-E_F|/V_{\rm min}$ ranges from 0.86 to 2.0 depending on the asymmetry of the molecular junction. In
order to use TVS as a quantitative tool, knowledge of the asymmetry
factor is therefore needed. On the other hand, if the molecular levels can be
determined by other means, the TVS can by used to measure the asymmetry factor.

\begin{figure}[htb!]
\includegraphics[width=0.9\columnwidth]{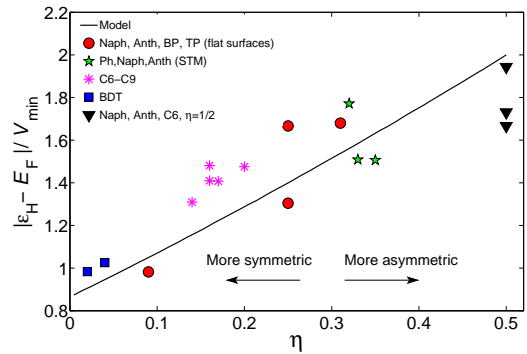}
    \caption{Ratio between the HOMO energy (at zero bias) and the transition voltage,
$V_{\rm min}$, vs. asymmetry parameter, $\eta$. Solid line is obtained from a
Lorentzian transmission function using Eqs. \eqref{Transmission} and
\eqref{Current} and symbols are results of \textit{ab-initio} finite bias
calculations.}
\label{TVS_data}
\end{figure}

\begin{figure}[htb!]
\begin{center}
 \includegraphics[width=0.4\columnwidth]{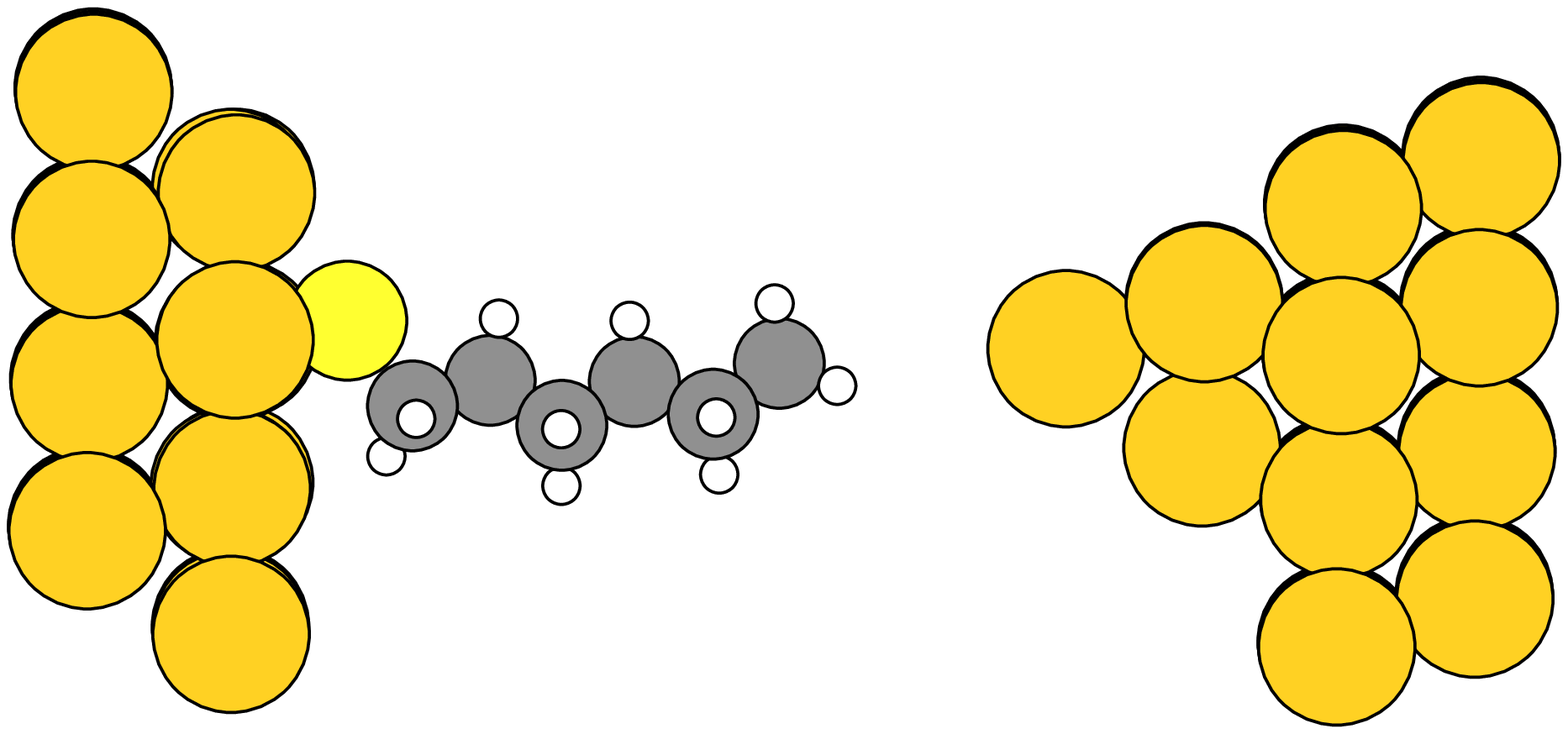}
\includegraphics[width=0.4\columnwidth]{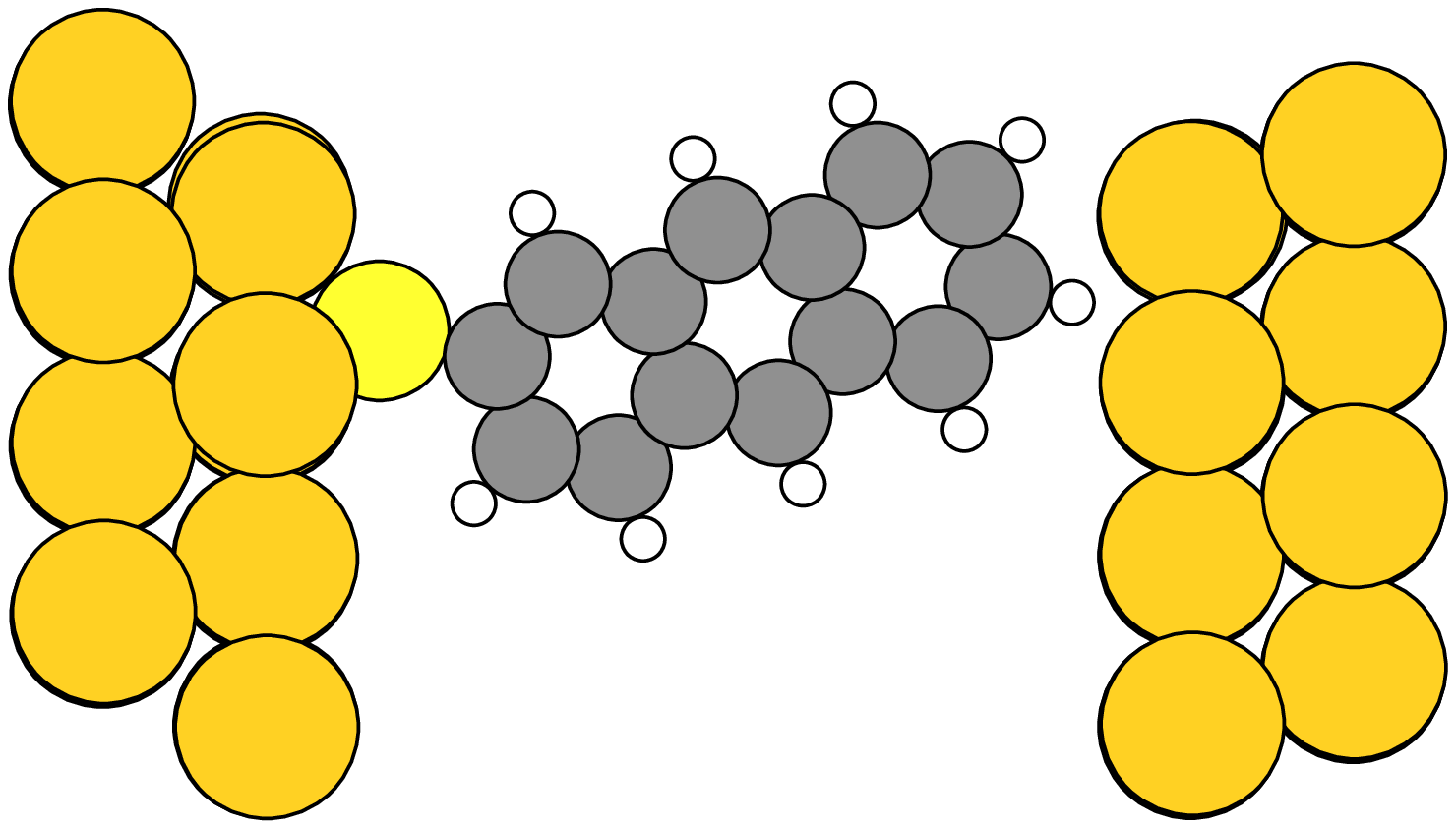}
\includegraphics[width=.9\columnwidth]{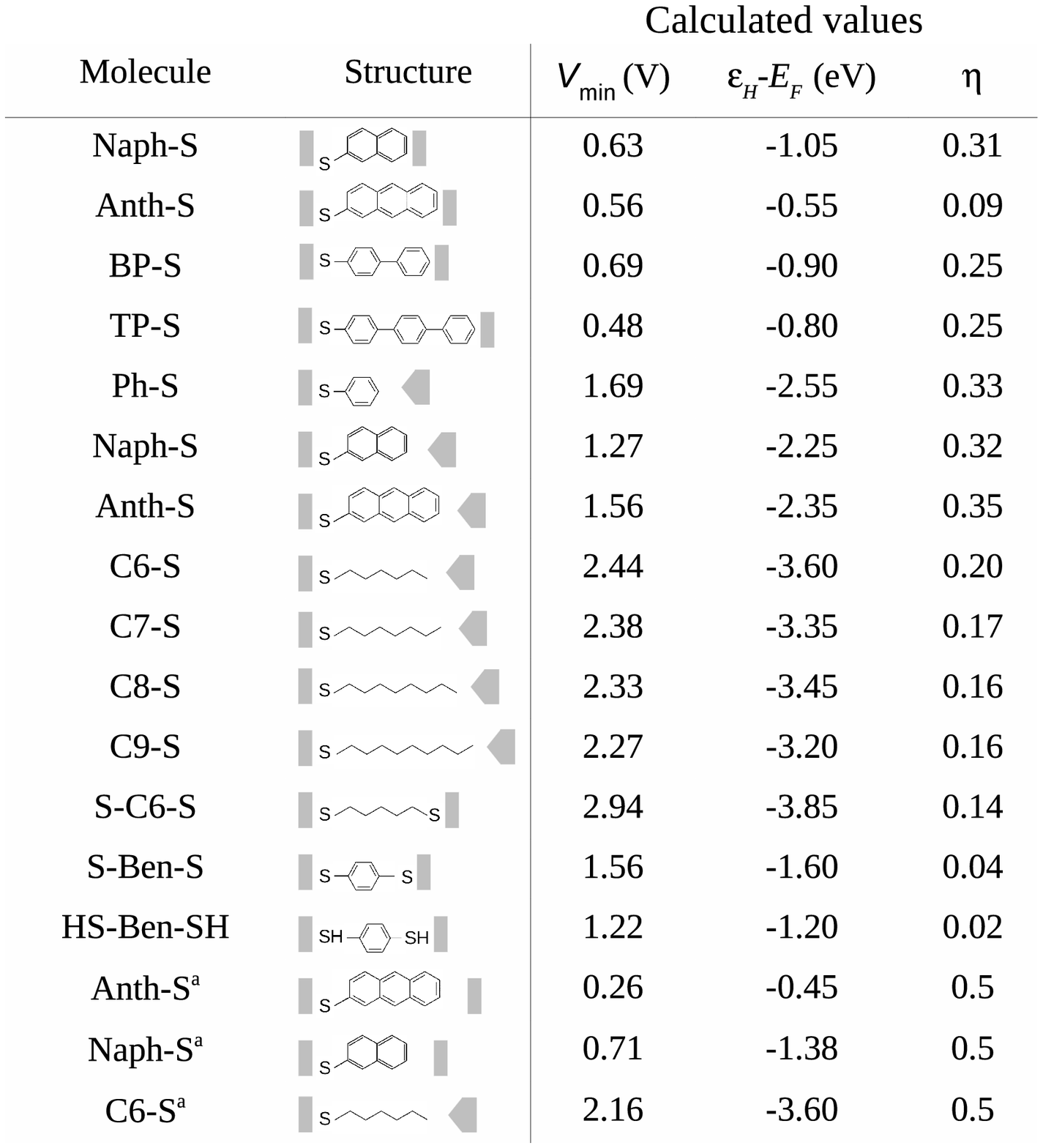}
\end{center}
    \caption{Top: Atomic structure of two of the major groups of molecular transport
junctions investigated in this work and in experiments~\cite{Beebe2006,Beebe2008,SongNature2009}, namely alkanes in an STM-like
configuration (top left), and anthracenethiol (Anth) between two flat Au (111)
surfaces (top right). The table in the lower part shows schematically all the
considered molecular junctions together with calculated values of $V_{min}$, $\varepsilon_H-E_F$, and $\eta$. The electrode configurations are indicated with
gray boxes showing both the flat Au surface (rectangles) and STM-like geometries
(pentagons). The super-script $^a$ indicates that the data are obtained from
non-self consistent calculations by manually setting $\eta=1/2$. The HOMO energy, $\varepsilon_H$, refers to the first peak in 
the Kohn-Sham (PBE) projected density of states. See text for more
details.}
\label{table-fig}
\end{figure}

\begin{figure}[htb!]
\includegraphics[width=0.8\columnwidth]{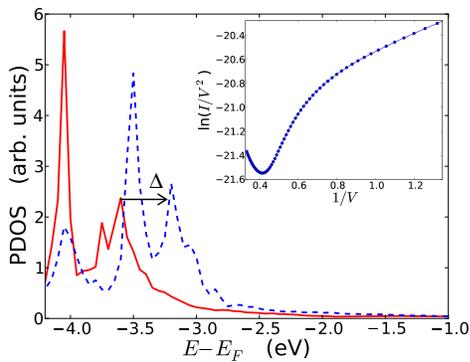}
\caption{Projected density of states on the C-atoms in the
  C$_6$-alkane junction shown in Fig. \ref{table-fig}. The solid red
  line refers to $V_{\rm bias}=0\,$V and the dashed blue line refers to
  $V_{\rm bias}=2.0\,$V. The HOMO level (at $V_{\rm bias}=0\,$V) is located at
  $E-E_F=-3.6\,$eV. We determine the asymmetry factor, $\eta$, from
  the shift, $\Delta$, of the HOMO level with the bias voltage. The inset
  shows the calculated Fowler-Nordheim plot with the characteristic
  TVS-minimum at $V_{\rm min}=2.4\,$V.}
\label{TVS_pdos_C6}
\end{figure}

We now turn to our {\it ab-initio} finite bias calculations and Fig.
\ref{TVS_data}, which is our main result. It shows the ratio of
$|\varepsilon_{\text{H}}-E_F|/V_{\rm min}$ vs. $\eta$ for the 17
molecular junctions listed in Fig.  \ref{table-fig}. The solid line is
the result obtained from the one-level model using Eqs.
\eqref{Transmission} and \eqref{Current}. Although there are
deviations between the model and the \textit{ab-initio} results, both
data sets clearly follow the same trend. This result further supports
TVS as a spectroscopic tool, but it also underlines that quantitative
information about the molecular level position can only be obtained
from TVS provided some knowledge of the junction asymmetry. The
calculated numbers for the data points are listed in Fig.
\ref{table-fig}.

The current at finite bias voltage is calculated using DFT in
combination with a non-equilibrium Green function (NEGF) method,
following the principles of Ref.~\cite{Brandbyge2002}. Our DFT-NEGF
method is implemented in GPAW, which is a real space electronic
structure code based on the projector augmented wave
method~\cite{gpaw,gpaw-lcao}. We use the PBE exchange-correlation
functional~\cite{PBE}, and a $4\times4$ $k$-point sampling in the
surface plane. The electronic wave functions are expanded in an atomic
orbital basis~\cite{gpaw-lcao}. In all calculations, the molecule and
the closest Au layers are described by a double-zeta plus polarization
(dzp) basis set, while the remaining Au atoms are described by a
single-zeta plus polarization (szp) basis.

The upper panel of Fig. \ref{table-fig} shows the atomic structure of
two representative junctions: a hexanethiol (left) in an STM-like
configuration with the STM-tip $\sim 4\,$\AA~away from the molecule,
and anthracenethiol (right) between two flat Au (111) surfaces with a
distance of 1.5 \AA~between the right Au
electrode and the closest H atom of the molecule. For all STM-setups, we initially relax the
molecule on a single Au(111) surface and subsequently add the STM
electrode without further relaxations.  For the systems with two flat
Au surfaces, we relax the molecule and the two closest Au layers.
Except for the benzene junctions (S-Ben-S and HS-Ben-SH), the linking
S atom relaxes into a bridge site of the Au(111) surface shifted
slightly towards the hollow site. The structure of the symmetric
benzene-dithiol (S-Ben-S) junction is taken from
Ref.~\cite{Strange2010} where the molecule binds to an Au adatom
together with a SCH$_3$ unit. In the other benzene junction
(HS-Ben-SH) the hydrogenated sulfur atoms bind to Au adatoms of the
Au(111) surfaces following Ref.~\cite{Ning2010}.

In Fig. \ref{table-fig} we also list for each junction the obtained
TVS minimum voltage, $V_{\rm min}$, the position of the HOMO level
relative to $E_F$, and the calculated asymmetry parameter. Figure
\ref{TVS_pdos_C6} illustrates how we determine these quantities, in
the case of the C6-S junction. The TVS minimum is simply found from
the minimum in the Fowler-Nordheim plot, as show in the inset. To
determine the HOMO level, we project the density of states onto the
carbon atoms. Molecular levels then appears as clear peaks in the projected density of state (PDOS)
vs. energy plot as shown in Fig.~\ref{TVS_pdos_C6}. We note that there
are situations where a peak in the PDOS does not show in the
transmission function due to symmetry mismatch between the molecular
orbital and the electrode wavefunctions. This is
indeed the situation for Ph-S, Naph-S and Anth-S in the STM setup where a peak in the PDOS is
observed at $E-E_F\sim -0.55$eV, but only a vanishingly small transmission peak occurs at this energy. In these three cases we take the second PDOS peak at $E-E_F=-2.35$eV as the HOMO level. At this energy, the Au tip
wavefunctions also have $d$-character, which enables non-zero coupling to the molecular $\pi$-orbital. 

At finite bias voltage, the energy of the HOMO level is shifted by the bias voltage. (c.f. Fig.
\ref{tvs_analytical}). Figure \ref{TVS_pdos_C6} shows the PDOS of the C$_6$-alkane junction
under bias voltage $V=0\,$V (red solid) and $V=2.0\,$V (blue dashed). The equilibrium HOMO energy of 
$\varepsilon_H-E_F=-3.6\,$eV shifts upward by $\Delta=0.4\,$eV when the bias is applied.
The asymmetry factor, $\eta$, is calculated from the shift, $\Delta$,
according to $\eta=\Delta/V=0.2$. 

Notice that even for the very asymmetric STM configurations, the
asymmetry factor is never larger than 0.35. This is due to the
relatively small distance between the molecule and the Au tip (4.0
\AA). Increasing this distance would increase $\eta$ further, however,
this is computationally problematic due to the finite range of the
atomic orbital basis. Instead we simulate a completely asymmetric
junction ($\eta=0.5$), by using the zero-bias
transmission function and fixing the electrode chemical potentials to
$\mu_L=E_F$ and $\mu_R=E_F-V$.
The data for the three last junctions
in Fig. \ref{table-fig} are obtained in this way and are thus non-self
consistent results.

The deviations of the numerical data from the analytical result in
Fig. \ref{TVS_data} may be due to several reasons. (i): The
\textit{ab-initio} transmission functions do not have a perfect
Lorentzian shape as assumed in the analytical model. (ii) The
dependence of the HOMO energy on the bias voltage is not strictly
linear. (iii) The bias voltage influences the shape of the molecular
orbitals which in turn affects the coupling strength. (iv) It may in
some cases be difficult to accurately determine $\varepsilon_H$ and
$\eta$, as discussed above.

The TVS minimum voltage was originally interpreted as a transition from tunneling to field emission~\cite{Beebe2006}. However, according to our calculations, $V_{\rm min}$ does not mark such a transition, and in line with Refs. \cite{Huisman2009,AraidaiPRB2010} we do not view $V_{\rm min}$ as a special transition voltage. Rather the usefulness of TVS relies on the direct relation between $V_{\rm min}$ and $\varepsilon_H$ when the transmission function can be well described by a Lorentzian. 

In the conducting AFM measurements of Ref.~\onlinecite{Beebe2006}, the
ratio $|\varepsilon_H-E_F|/V_{\rm min}$ was found to vary between 2.1
and 2.6 for a set of molecules which included Naph-S, BP-S, and TP-S
also considered in this work. This is significantly larger than the
value of 1 predicted by Simmons barrier model. According to our model
the ratio should be smaller than 2.0 which is obtained for a
completely asymmetric junction ($\eta=0.5$). To resolve this puzzle,
we first note that our \emph{ab-initio} calculations yield
$\eta\sim0.3$ for the three molecules (c.f. Fig. \ref{table-fig}) giving
$|\varepsilon_H-E_F|/V_{\rm min}\sim 1.6$ according to our model.
Next, we note that the presence of the AFM tip in the transport
measurements will lead to a renormalization of the HOMO energy due to
an image charge effect~\cite{juanma09} which is not present in the UPS
measurements. Based on the method described in Ref.
\onlinecite{duncan} we estimate the image charge interaction to be
$0.5-0.35\,$eV depending on the size of the molecule. Correcting the UPS
values for $\varepsilon_H$ by these values lead to ratios
$|\varepsilon_H-E_F|/V_{\rm min}$ in the range $1.6-1.8$ in good
agreement with the model prediction of $\sim 1.6$.

Finally, we note that the well known inability of DFT to describe 
energy gaps and level alignment of molecules at
surfaces~\cite{juanma09} does not affect the conclusions of the present work.
This is because, according to Fig.  \ref{tvs_analytical}, the
\emph{ratio} $(\varepsilon_H-E_F)/V_{\rm min}$ is independent of the
value of $\varepsilon_H$. In particular, the dependence of
$(\varepsilon_H-E_F)/V_{\rm min}$ on asymmetry is expected to be a
general result independent of the absolute position of the molecular
levels.

In conclusion, we have performed extensive \textit{ab initio} DFT transport
calculations to simulate transition voltage spectroscopy for a large number of
molecular junctions. The numerical data closely follow the trend expected from an
analytical model with a Lorentzian shaped transmission function. We have explicitly shown
that in order to use TVS as a quantitative spectroscopic tool to probe the molecular
levels, it is necessary to take the asymmetry of the molecular junction into
account. The present analysis should therefore be considered in future applications of
transition voltage spectroscopy.

\begin{acknowledgements}
The center for Atomic-scale Materials Design (CAMD) is 
funded by the Lundbeck Foundation. The authors acknowledge
support from FTP through grant no. 274-08-0408 and from 
The Danish Center for Scientific Computing. 

\end{acknowledgements}


\end{document}